\documentclass[aps,twocolumn,superscriptaddress]{revtex4-2}
\usepackage{color}
\usepackage{amsmath}
\usepackage{graphicx}
\usepackage{bm}
\usepackage{braket}
\usepackage{colortbl}
\usepackage{comment}
\usepackage{graphicx}
\usepackage{color}
\usepackage[]{natbib}
\usepackage{hyperref}
\usepackage{url}
\usepackage{here}

\begin{abstract}
Quantum dots are recognized as a suitable platform for studying thermodynamic phenomena involving single electronic charges and spins in nano-scale devices. However, such a thermodynamic system is usually driven by electron reservoirs at different temperatures, not by a lattice temperature gradient. We report on experimental observations of charge-spin cooperative dynamics in transitions of two-electron spin states in a GaAs double quantum dot located in a non-equilibrium phonon environment. Enhancements in the spin-flip processes are observed, originating from phonon excitation combined with the spin-orbit interaction. In addition, due to the spatial gradient of phonon density between the dots, the spin-flip rate during an inter-dot electron tunnel from a hot to a cold dot is more enhanced than in the other direction, resulting in accumulation of parallel spin states in the double dot.
\end{abstract}

\begin{document}
\title{Real-time Observation of Charge-spin Cooperative Dynamics\\ Driven by a Nonequilibrium Phonon Environment}
\author{Kazuyuki Kuroyama}
\email[]{kuroyama@iis.u-tokyo.ac.jp}
\affiliation{Department of Applied Physics, The University of Tokyo,
7-3-1 Hongo, Tokyo 113-8656, Japan}
\author{Sadashige Matsuo}
\email[]{sadashige.matsuo@riken.jp}
\affiliation{Department of Applied Physics, The University of Tokyo,
7-3-1 Hongo, Tokyo 113-8656, Japan}
\affiliation{Center for Emergent Materials Science (CEMS), RIKEN, 2-1 Hirosawa,
Wako-shi, Saitama 351-0198, Japan}
\affiliation{Japan Science and Technology Agency, PRESTO, 4-1-8 Honcho, Kawaguchi-shi, Saitama 332-0012, Japan}
\author{Jo Muramoto}
\affiliation{Department of Applied Physics, The University of Tokyo,
7-3-1 Hongo, Tokyo 113-8656, Japan}
\author{Shunsuke Yabunaka}
\affiliation{Department of Physics, Kyushu University, Fukuoka 819-0395, Japan}
\author{Sasha R. Valentin}
\affiliation{Lehrstuhl f\"{u}r Angewandte Festk\"{o}rperphysik, Ruhr-Universit\"{a}t,
Bochum, Universit\"{a}tsstrasse 150, D-44780 Bochum, German}
\author{Arne Ludwig}
\affiliation{Lehrstuhl f\"{u}r Angewandte Festk\"{o}rperphysik, Ruhr-Universit\"{a}t,
Bochum, Universit\"{a}tsstrasse 150, D-44780 Bochum, German}
\author{Andreas D. Wieck}
\affiliation{Lehrstuhl f\"{u}r Angewandte Festk\"{o}rperphysik, Ruhr-Universit\"{a}t,
Bochum, Universit\"{a}tsstrasse 150, D-44780 Bochum, German}
\author{Yasuhiro Tokura}
\affiliation{Pure and Applied Sciences, University of Tsukuba, 1-1-1 Tennodai, Tsukuba, Ibaraki 305-8571, Japan}
\author{Seigo Tarucha}
\email[]{tarucha@riken.jp}
\affiliation{Department of Applied Physics, The University of Tokyo,
7-3-1 Hongo, Tokyo 113-8656, Japan}
\affiliation{Center for Emergent Materials Science (CEMS), RIKEN, 2-1 Hirosawa,
Wako-shi, Saitama 351-0198, Japan}

\maketitle

\section{Introduction}
A heat engine consisting of a single electron and its spin has recently attracted increased attention from the  perspectives of energy harvesting and thermoelectric conversion for waste heat created in nano-scale electric devices \cite{Roche2015,PhysRevLett.123.117701}. In general, according to the second law of thermodynamics, to drive a thermodynamic cycle in such a heat engine and extract work, two heat reservoirs at different temperatures are necessary, meaning that  a thermodynamic device must be located in the non-equilibrium environment of heat. For the demonstration of a heat engine in nano devices, a quantum dot (QD) is recognized as one of the best systems due to its high controllability of both single electron charges and their spins, and a number of related studies on QD-based heat engines have been reported to date \cite{Thierschmann2015,Josefsson2018, PhysRevLett.123.117701,PhysRevLett.125.166802,PhysRevB.99.235432}. However, in these studies, a QD heat engine is driven by electron reservoirs at different \textit{electron} temperatures, not by thermal reservoirs at different \textit{lattice} temperatures. This is probably because creating a lattice temperature gradient over a distance of at most a few hundred nanometer, the order of QD size, and controlling the lattice temperature without disturbing the electron temperature in such a nanoscale system is technically challenging, although a QD-based local phonon source can be implemented in a lateral QD device \cite{PhysRevLett.100.176805, PhysRevLett.104.196801, Granger2012}. Such a heat engine driven by a local lattice temperature gradient would aid understanding of thermodynamic and thermoelectric phenomena of electrons in mesoscopic systems and, regarding practical applications, for improving the coherence time of solid-state quantum bits, like spin and charge qubits with QDs, that are sensitive to the phonon environment \cite{PhysRevLett.98.126601,PhysRevB.89.085410,Purkayastha2020}.

In this work, we concentrate on the charge-spin cooperative dynamics of two-electron states in a GaAs lateral double QD (DQD) \cite{RevModPhys.79.1217} driven by a nonequilibrium phonon environment created by a QD-based phonon source located at one side of the DQD. As an electron feels a different phonon density in each dot, the charge-spin dynamics reflect the spatial gradient of a phonon density. To observe the phonon-induced charge-spin dynamics at a single electron level, we use a real-time charge sensing technique with a nearby QD \cite{Lu2003Nature, doi:10.1063/1.1815041, doi:10.1063/1.1784875}. A generated phonon excites an electron in the DQD to the first excited states, but the intra-dot excitation and relaxation are not observable by charge sensing because of no charge transfer occurring. However, as the intra-dot phonon excitation needs a spin flip due to the selection rule of an orbital and spin angular momentums, such an excitation can be observed as its spin-flip of an electron. Based on this fact, we introduce the Pauli spin-blockade effect (PSB) of a DQD to observe the phonon-induced spin-flip events \cite{Ono1313, PhysRevB.72.165308}. In the case of a DQD holding two or more electrons, electron tunneling between the dots is strongly affected by their spin configurations, due to this spin-blockade effect \cite{PhysRevLett.116.136803, PhysRevLett.117.206802,PhysRevLett.119.176807}. Thereby, we count the number of the spin-flip events in the two-electron DQD set in the PSB regime under phonon irradiation and thus to obtain the full counting statistics (FCS) of the charge-spin dynamics \cite{PhysRevResearch.2.033120}. Our results show that the spin-flip rate increases notably when the generated single phonon energy exceeds the lowest excitation energy in the DQD. To evaluate the phonon-induced transition rates, we extend an ordinary FCS technique to include not only the ground state but also the excited states. The theoretical calculation considering both the spin-orbit and electron-phonon interactions shows that an inter-dot electron tunnel through the excited states results in an increased spin-flip rate. Finally, we discover that the spin-flip tunnel process depends on the inter-dot tunneling direction under phonon irradiation. This result implies that the phonon density has a spatial gradient between two dots of the DQD, and accordingly, the spin-flip tunneling processes are strongly modulated by the nonequilibrium phonon environment.  

\section{Results}
\begin{figure*}[hbt]
\centering
\includegraphics[width=\linewidth]{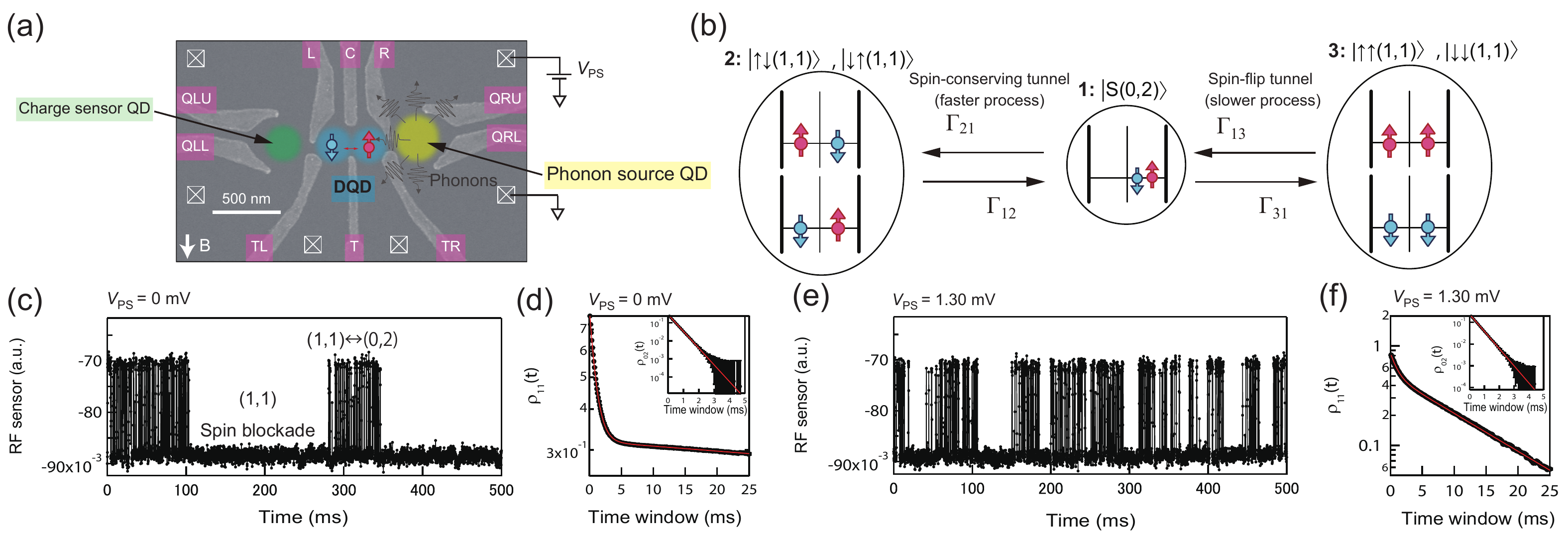} 
\caption{(a) Scanning electron micrograph of our sample.
The DQD is formed on the two blue circles.
The TL, T, and TR gate voltages are tuned such that the trapping time of
an electron is sufficiently longer than the spin-flip time, a few hundred
milliseconds. The charge sensor is placed on the left green circle.
The phonon source is located on the right yellow circle, on which the DC bias
voltage $V_{\mathrm{PS}}$ is applied. (b) Transition diagram in a resonant two-electron DQD. States 1 and 2 (1 and 3) are connected by the spin-conserving (spin-flip) tunnels. (c) Time trace on the (0,2)-(1,1) resonance condition with no phonon irradiation. From 0 to 100 ms and from 285 to 350 ms, the charge state repeatedly and telegraphically oscillates between the two levels referring to the (0,2) and (1,1) charge states. Around 100 ms, the state starts to be blocked in the (1,1) states until 285 ms, indicating that the spin state is the $\ket{\uparrow\uparrow(1,1)}$ or $\ket{\downarrow\downarrow(1,1)}$ state after an accidental spin-flip tunnel process. (d) FCS distribution regarding the (1,1) [$\rho_{11}(0,t)$] at zero bias voltage $V_{\mathrm{PS}}$. The numerical fitting result is colored red. The loose slope is assigned to the spin-flip tunnel rate. The inset is the FCS distribution of the (0,2) charge state, $\rho_{02}(0,t)$. (e) Time trace with phonon irradiation at $V_{\mathrm{PS}}$ = 1.30 mV. Both the oscillating and blocked times are shorter than those in Fig. \ref{fig1}(c), indicating that the spin-flip rate increases. (f) $\rho_{11}(0,t)$ constructed at $V_{\mathrm{PS}}$ = 1.30 mV. As expected from Figs. \ref{fig1}(c) and (e), the loose slope becomes steeper. The inset shows $\rho_{02}(0,t)$.}
\label{fig1} 
\end{figure*}
The gate electrode configuration of our DQD devices fabricated from a GaAs quantum well wafer is depicted in Fig. \ref{fig1}(a). 
The DQD potential is defined by applying negative voltages on the gate electrodes TL, T, TR, L, C, and R. The Ohmic contacts are indicated by the white crosses. The charge states of the DQD are monitored by a nearby QD charge sensor located on the left side, which is embedded in a radio frequency (RF) resonant electrical circuit. An RF voltage signal with a frequency of 285.5 MHz is continuously applied to the sensor through the RF resonant circuit to perform RF reflectometry of the electrical conductance at the sensor \cite{doi:10.1063/1.2794995,Barthel2010Phys.Rev.B}. We note that no bias voltage is applied on the charge sensor to obtain the highest sensitivity, and thus, no phonon emission is expected to originate from the sensor QD.  

To generate nonequilibrium phonons, we separately install an additional QD formed right next to the DQD as shown in yellow [see Fig. \ref{fig1}(a)] \cite{Cao_2013,PhysRevB.79.035303,PhysRevLett.104.196801,Granger2012}. 
A relatively large DC bias voltage $V_{\mathrm{PS}}$ is applied on the QD, injecting hot electrons, which accompanies phonon emissions through the inelastic relaxation process. Therefore, this QD is regarded as a phonon source. We note that the highest phonon energy emitted from the source is  $eV_{\mathrm{PS}}$ \cite{PhysRevLett.104.196801}.

We set the gate voltages, L and R, such that the chemical potentials of the two dots are equivalent between the lowest (1,1) and (0,2) charge states, where the integers $i,j$ in the brackets denote the electron occupations of the left and right dot, respectively [see Fig. \ref{fig1}(b)]. The lowest (0,2) state is a singlet ($\ket{S(0,2)}=(\ket{\uparrow\downarrow(0,2)}-\ket{\downarrow\uparrow(0,2)})/\sqrt{2}$). The inter-dot coupling of the order of 1 neV is tuned to be much smaller than the local Zeeman energy difference $E_{Z}$ (of the order of 10 neV) between the two dots, which arises from the random fluctuation of a surrounding nuclear field of a few millitesla, originating from Ga and As atoms \cite{RevModPhys.79.1217}. Hence, for the (1,1) states, the singlet $\ket{S(1,1)}$ and one of the triplet states, $\ket{T_0(1,1)}$, are no longer eigenstates, whereas $\ket{\uparrow\downarrow(1,1)}$ and $\ket{\downarrow\uparrow(1,1)}$ are. To turn on the PSB, we apply an in-plane magnetic field of 100 mT which is much larger than the fluctuating nuclear field. For states with parallel spins, inter-dot electron tunneling from (1,1) to (0,2) is prohibited because the (0,2) state with parallel spins ($\ket{T_{\pm}(0,2)}$) is energetically much higher than $\ket{S(0,2)}$. Note that the energy separation between $\ket{T_{\pm}(0,2)}$ and $\ket{S(0,2)}$ is approximately 1 meV, much larger than any other energy scales of interest here. Therefore, the inter-dot electron tunneling only occurs when accompanied by a spin flip to break the PSB. In contrast, when two spins are antiparallel, the electron can resonantly tunnel back and forth between the two dots through the singlet component of the $\ket{\uparrow\downarrow(1,1)}$ or $\ket{\downarrow\uparrow(1,1)}$ states. Finally, we should note that although $\ket{\uparrow\uparrow(1,1)}$ and $\ket{\downarrow\downarrow(1,1)}$ are separated from the anti-parallel spin states by the Zeeman effect, as the electron temperature is approximately 100 mK, the thermal energy should be comparable with the Zeeman energy. Therefore, all four (1,1) states can be accessed with mostly equal probability. 

There are two possible spin-flip mechanism in electron tunneling through a coupled QD: spin-orbit interaction during the inter-dot charge tunnel \cite{PhysRevLett.116.136803, PhysRevLett.117.206802} and hyperfine interaction in each dot \cite{PhysRevLett.88.186802}. A number of previous studies have already shown that the spin-orbit interaction is more dominant in our experimental condition (see section VII in Supplemental Material). Therefore, we only consider the spin-orbit interaction in the following discussions, and then assume a transition diagram of the two-electron spin states as shown in Fig. \ref{fig1}(b). In this figure, $\Gamma_{ij}$ indicates the tunnel rate in the transition from the electron state $j$ to $i$. $\Gamma_{21}$ and $\Gamma_{12}$ are the spin-conserving tunnel rates, while $\Gamma_{31}$ and $\Gamma_{13}$ are the spin-flip tunnel rates. We note that the spin-flip tunneling is usually much slower than the spin-conserving tunneling.

First, we measure the inter-dot resonant tunneling in the PSB regime with no phonon irradiation, i.e., $V_{\mathrm{PS}}=0$ mV.
Figure \ref{fig1}(c) indicates an obtained typical time trace of the RF charge sensing signal. The time trace shows a two-level telegraph signal, indicating that the DQD charge state is either (0,2) or (1,1). Between 0 and 100 ms, and between 285 and 350 ms, fast inter-dot transitions between the (0,2) and (1,1) charge states are observed iteratively, implying that the spin configuration is antiparallel [see Fig. \ref{fig1}(b)].  
In contrast, the stable region, in which the charge state stays at (1,1) for a long time from 100 to 285 ms and from 350 to 500 ms, appears due to the prohibition of the charge transition by the PSB effect. Therefore, the spin configuration in this blockade region is supposed to be parallel, either $\ket{\uparrow\uparrow(1,1)}$ or $\ket{\downarrow\downarrow(1,1)}$.

We use the FCS method to evaluate the spin-conserving and spin-flip tunnel rates from the experimental data, considering a probability of $n$ instances of inter-dot tunneling for a certain time window $t$ and a final charge state of either (1,1) or (0,2). In the following discussion, we focus on only the probability of $n=0$, which is the same as that of state (1,1) [or (0,2)] without any inter-dot tunneling in the time window $t$. In this situation, we omit the notation $n$ for simplicity. The FCS probability distributions of the (1,1) charge state [$\equiv\mathrm{\rho_{11}}(n=0,t)=\mathrm{\rho_{11}}(t)$)] and the (0,2) state [$\equiv\mathrm{\rho_{02}}(n=0,t)=\mathrm{\rho_{02}}(t)$] are analytically derived as follows.
\begin{align}
\mathrm{\rho_{11}}(t)&=C_2\mathrm{e}^{-\Gamma_{12}t}+C_3\mathrm{e}^{-\Gamma_{13}t},\label{DoubleEXP}\\
\mathrm{\rho_{02}}(t)&=C_1\mathrm{e}^{-(\Gamma_{21}+\Gamma_{31})t},
\label{SingleEXP}
\end{align}
where the coefficient of $C_i$ ($i=$ 1, 2, 3) is the population at state $i$ depicted in Fig. \ref{fig1}(b). These coefficients are represented by the ratios of the transition rates as explained in section VI of the Supplementary Material. Figure \ref{fig1}(d) depicts the probability distributions $\mathrm{\rho_{11}}(t)$ (inset: $\mathrm{\rho_{02}}(t)$) constructed from the measured time traces. For the $\mathrm{\rho_{11}}(t)$, we clearly observe a steep and loose slope. These two slopes are assigned to the spin-conserving ($\Gamma_{12}$) and spin-flip ($\Gamma_{13}$) tunnel processes, corresponding to the $\mathrm{e}^{-\Gamma_{12}t}$ and $\mathrm{e}^{-\Gamma_{13}t}$ terms in Eq. (\ref{DoubleEXP}), respectively.
The red curve is the fitted function of Eq. (\ref{DoubleEXP}) [inset: Eq. (\ref{SingleEXP})] to the experimental data. 
From this fitting and using our compensation technique discussed in section V of Supplemental Material, $\Gamma_{12}$ of 1.21 kHz and $\Gamma_{13}$ of 3.51 Hz are obtained. The remaining transition rates of $\Gamma_{21}$ = 2.32 kHz and $\Gamma_{31}$ = 4.80 Hz are evaluated, using the coefficients and exponent of $\mathrm{\rho_{02}}(t)$. 

Subsequently, we turn on the phonon source by applying a finite bias voltage $V_{\mathrm{PS}}$. Figure \ref{fig1}(e) is a typical time trace measured for $V_{\mathrm{PS}}=$ 1.30 mV. Compared to the result at $V_{\mathrm{PS}}=0$ mV in Fig. \ref{fig1} (c), the blockade times at the (1,1) state are shorter. This shorter blockade time can be interpreted as the spin-flip tunnel processes occurring more frequently under phonon irradiation. For quantitative comparison, the FCS probability distributions measured at $V_{\mathrm{PS}}=1.30$ mV are indicated in Fig. \ref{fig1} (f). We again see a definite feature of a double-exponential function in the $\mathrm{\rho_{11}}(t)$ distribution, but only the second slope reflecting the spin-flip tunnel rate becomes steeper as a larger bias voltage is applied. Using the same analysis as for Fig. \ref{fig1}(c), $\Gamma_{12}$ of 1.06 kHz, $\Gamma_{21}$ of 1.70 kHz, $\Gamma_{13}$ of 88.9 Hz, and $\Gamma_{31}$ of 170 Hz are obtained. Thus, we confirm that the spin-flip tunnel rate ($\Gamma_{13}$, $\Gamma_{31}$) increases more than tenfold by phonon irradiation. 
\begin{figure}[h]
\centering{ \includegraphics[width=\linewidth]{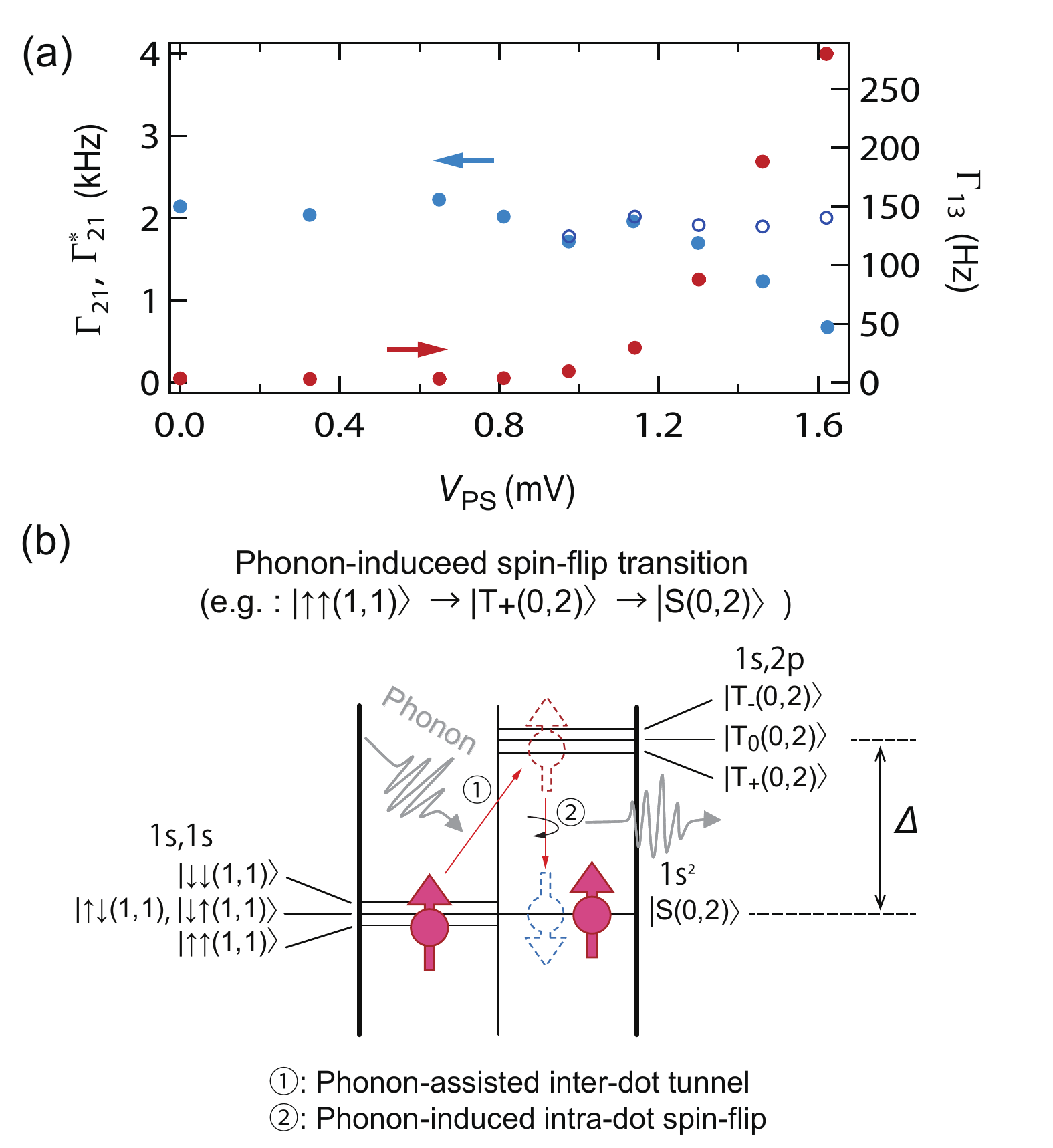} }
\caption{(a) Spin-conserving tunnel rate $\Gamma_{21}$ (filled blue circles) and spin-flip tunnel rate $\Gamma_{13}$ (red circles) with respect to the bias voltage of the phonon source $V_{\mathrm{PS}}$. The enhancement is only observed in $\Gamma_{13}$ not in $\Gamma_{21}$. We additionally plot $\Gamma_{21}^*$ evaluated by the five-state FCS calculation with open deep-blue circles. (b) Energy level diagram in a DQD, which explains the phonon-induced spin-flip tunneling process. When the spin state is initially $\ket{\uparrow\uparrow(1,1)}$ ($\ket{\downarrow\downarrow(1,1)}$), a phonon with energy equal to $\Delta$ excites the electron in the left dot to the 2p-orbital of the right dot and forms $\ket{T_+(0,2)}$ ($\ket{T_-(0,2)}$). Subsequently, due to the combination of the spin-orbit and electron-phonon interactions, the electron spin in the 2p-orbital flips accompanied by the phonon relaxation down to $\ket{S(0,2)}$. The opposite spin-flip process, $\ket{S(0,2)}\rightarrow\ket{T_+(0,2)}\rightarrow\ket{\uparrow\uparrow(1,1)}$ (and $\ket{S(0,2)}\rightarrow\ket{T_-(0,2)}\rightarrow\ket{\downarrow\downarrow(1,1)}$), is also available.}
\label{fig2} 
\end{figure}

To reveal the dependence of the transition rates on the phonon energy, we evaluate $\Gamma_{21}$ and $\Gamma_{13}$ at various values of $V_{\mathrm{PS}}$ ranging from 0 to 1.60 mV. 
The obtained $\Gamma_{21}$ and $\Gamma_{13}$ are plotted by the blue and red closed circles in Fig. \ref{fig2}(a), respectively. Both $\Gamma_{21}$ and $\Gamma_{13}$ are unchanged for  $V_{\mathrm{PS}}<$  0.90 mV, but for further increasing $V_{\mathrm{PS}}$, $\Gamma_{13}$ significantly increases, whereas $\Gamma_{21}$ gradually decreases. As the spin-orbit effect is determined by the material and the relative orientation of the QD array to the crystallographic axis \cite{PhysRevB.88.075306} and the magnetic field direction (see sections II and VII in Supplemental Material), the ratio of $\Gamma_{13}/\Gamma_{21}$ is anticipated to be constant with phonon irradiation. 
Therefore, the obtained enhancement of $\Gamma_{13}$ can not be explained by the spin-orbit effect of the ground states, as discussed in previous studies \cite{PhysRevLett.116.136803,PhysRevLett.117.206802}. Moreover, phonons have only a little effect on the inter-dot transitions between the ground states and on the electron temperature, which is defined by the linewidth of the inter-dot resonant tunneling, as discussed in section IV in Supplemental Material.
 
As the enhancement of the spin-flip tunnel rate is observed only when the phonon energy exceeds the threshold voltage of approximately $V_{\mathrm{PS}}=$ 0.90 mV, it is reasonable to assume that the excitation process in the DQD plays an important role. 
The lowest excited states, $\ket{T_{\pm}(0,2)}$ and $\ket{T_{0}(0,2)}$, can be accessed from $\ket{S(0,2)}$ via phonon absorption and spin-orbit interaction, and also from the (1,1) states via phonon absorption [see Fig. \ref{fig2}(b)]. 
In order to confirm that the threshold of $V_{\mathrm{PS}}$ originates from the energy separation between $\ket{S(0,2)}$ and $\ket{T(0,2)}$ ($\equiv\Delta$), we analyze a charge stability diagram of the DQD (see section III in Supplemental Material). 
When the source-drain voltage is applied on the DQD, two sharp lines assigned to singlet and triplet resonances appear. 
From the separation between these two resonance lines, we can estimate the energy separation $\Delta$ = 0.95 meV (see section II in Supplemental Material). 
The value is within the typical range of 0.5-1 meV \cite{PhysRevB.70.241304,PhysRevLett.98.126601,PhysRevLett.115.176802} and also consistent with the threshold voltage of  $V_{\mathrm{PS}}$ = 0.90 mV. 

To explain the observed spin-flip rate enhancement, we propose its mechanism based on previous theoretical work \cite{PhysRevB.77.045328} and on the alignment of two-electron spin states as shown in Fig. \ref{fig2}(b). 
The increase in the spin-flip tunnel rates is explained by a combination of ``the phonon-assisted inter-dot transition with spin conservation between $\ket{T_{+}(0,2)}$ ($\ket{T_{-}(0,2)}$) and $\ket{\uparrow\uparrow(1,1)}$ ($\ket{\downarrow\downarrow(1,1)}$)" and ``the intra-dot spin-flip transition between $\ket{T_{\pm}(0,2)}$ and $\ket{S(0,2)}$". 
Here, we explain the transition process starting from the (1,1) charge state with parallel spins subject to the PSB. As the inter-dot coupling is sufficiently weak, an electron in each dot is located mainly in the s-type orbital (orbital angular momentum $l=0$). When phonon energy compensates the energy separation $\Delta$, an inter-dot transition from $\ket{\uparrow\uparrow(1,1)}$ ($\ket{\downarrow\downarrow(1,1)}$) to $\ket{T_{+}(0,2)}$ ($\ket{T_{-}(0,2)}$ ) is allowed. Although one of the electrons is excited between the orbitals with different angular momentum, i.e., from the s-type orbital ($l=0$) to the p-type orbital ($l=1$), this inter-dot transition is still allowed because of the lack of rotational symmetry of the DQD.
Subsequently, $\ket{T_{\pm}(0,2)}$ swiftly relaxes to $\ket{S(0,2)}$ by one phonon emission, mediated by the spin-flip induced by the spin-orbit interaction. In contrast to the phonon-assisted inter-dot transition, a spin is required to be flipped because the relaxation occurs inside the dot, which is mostly rotationally symmetric. This spin-flip relaxation process takes place within a few hundered milliseconds \cite{Fujisawa2002, PhysRevLett.95.056803, PhysRevLett.94.196802}, much faster than the other transitions, because of the larger dipole moment of the p-type orbital wave function \cite{PhysRevB.77.045328}.
\begin{figure}[h]
\centerline{ \includegraphics[width=\linewidth]{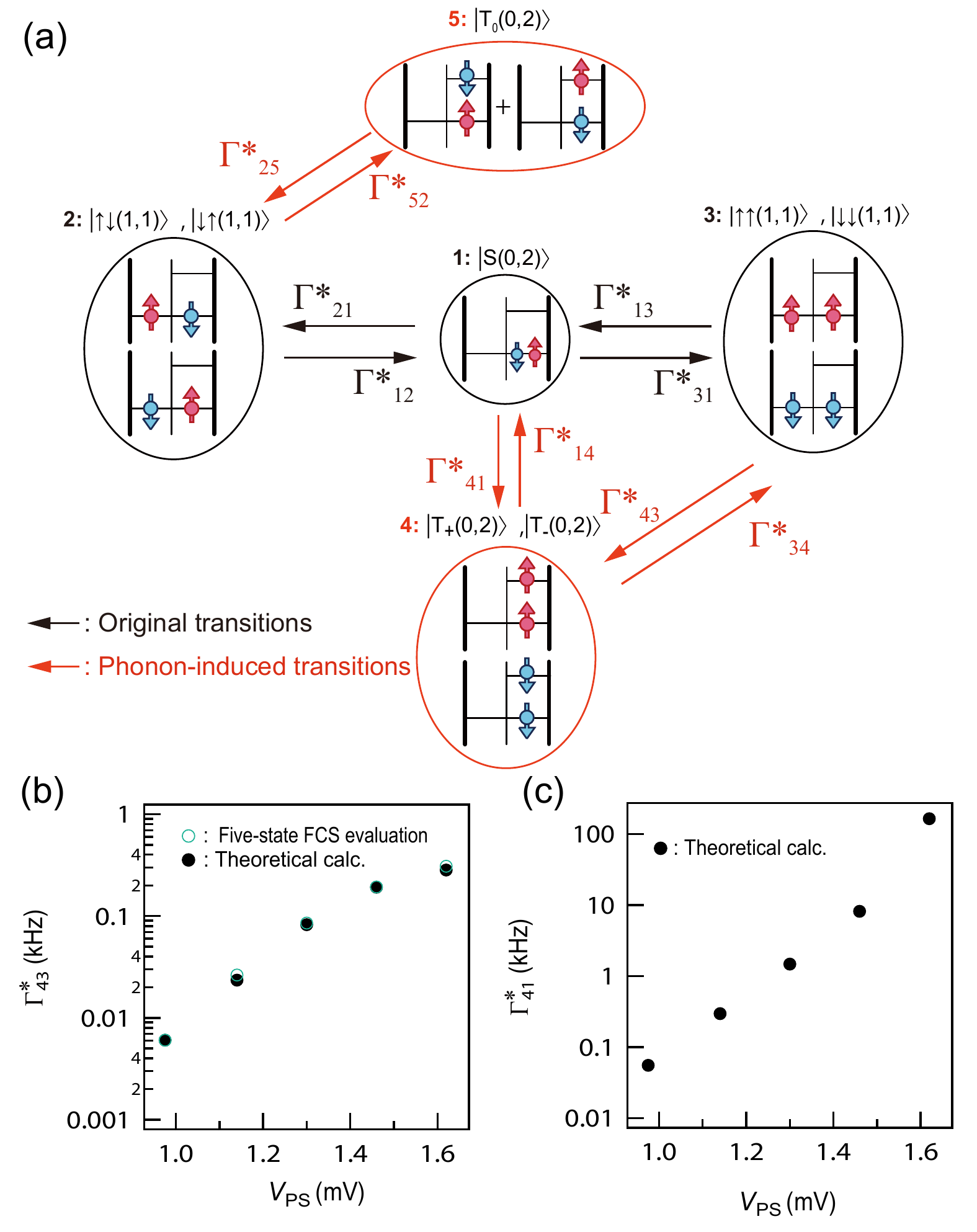} }
\caption{(a) Transition diagram used for the five-state FCS calculation. The newly added transitions are depicted by red arrows, and are induced by phonon irradiation. The transitions between states 3 and 4 and between 2 and 5 are phonon-assisted inter-dot tunneling without a spin flip, whereas those between states 1 and 4 are the phonon-induced spin-flip processes occurring in the right dot. (b) The phonon-assisted inter-dot tunnel rate $\Gamma_{43}^*$ of the FCS (green) and theoretical calculation (black) plotted with respect to $V_{\mathrm{PS}}$. (c) Theoretical calculation of the intra-dot spin-flip excitation rate $\Gamma_{41}^*$ using the same conditions as $\Gamma_{43}^*$.}
\label{fig3} 
\end{figure}

From the above discussion, we assign the first excited states of $\ket{T_{\pm}(0,2)}$ as responsible for the enhancement of the spin-flip tunnel rates. Therefore, we exploit our model by differentiating the (0,2) charge state into the ground state ($\ket{S(0,2)}$) and the three excited states ($\ket{T_{\pm}(0,2)},\ \ket{T_{0}(0,2)}$). The transition diagram is modified as depicted in Fig. \ref{fig3}(a). Both the spin-flip intra-dot and spin-conserving inter-dot transitions are mediated by a phonon process. Here, the transition rate $\Gamma_{ij}^*$ from state $j$ to $i$ is newly defined as the value evaluated using FCS for five states [Fig. \ref{fig3}(a)], to avoid confusion with the rates $\Gamma_{ij}$ evaluated used FCS for three states [Fig. \ref{fig1}(b)]. The excited state 4 ($\ket{T_{\pm}(0,2)}$) is accessed from state 3 with the phonon-assisted, spin-conserving inter-dot tunneling and also from state 1 with the intra-dot spin-flip process. It should be noted that the phonon-assisted inter-dot transitions between $\ket{T_{0}(0,2)}$ and $\ket{\uparrow\downarrow(1,1)}$ (or $\ket{\uparrow\downarrow(1,1)}$) are also available. However, the transition between $\ket{T_{0}(0,2)}$ and $\ket{S(0,2)}$ may be very slow because of the selection rule of the total angular momentum conservation in the spin-orbit interaction \cite{PhysRevLett.95.056803, PhysRevLett.98.126601}. Therefore, we dismiss this transition in the FCS computation. We derive a set of FCS differential equations (see section VI in Supplemental Material) and obtain the following FCS distributions for the (1,1) and (0,2) charge states.  
\begin{align}
\rho_{11}^*(t)&=C_2\mathrm{e}^{-(\Gamma_{12}^*+\Gamma_{52}^*)t}+C_3\mathrm{e}^{-(\Gamma_{13}^*+\Gamma_{43}^*)t},\\
\rho_{02}^*(t)&=(C_1^*+C_4^*)\mathrm{e}^{-\lambda_{-}^*t}+C_5^*\mathrm{e}^{-\Gamma_{25}^*t}.
\end{align}
The coefficient of $C_i^*$ ($i=$ 1, 4, 5) is the population of state $i$, which is one of the (0,2) charge states, and the relation $C_1^*+C_4^*+C_5^*=C_1$ is satisfied. $\lambda_{-}^*$ is no longer expressed in a simple manner like $\rho_{02}(t)$ in Eq. (\ref{SingleEXP}). We explain how to derive $\lambda_{-}^*$ in section VI of Supplemental Material. Similar to the three-state FCS, the distribution $\rho_{11}^*(t)$ consists of a double exponential function, and it clearly shows that the increase in the second slope of $\rho_{11}^*(t)$ originates from the phonon-assisted inter-dot tunnel rate $\Gamma_{43}^*$, whereas $\Gamma_{13}^*$ (and $\Gamma_{31}^*$) may be much less affected by phonons. For $\rho_{02}^*(t)$, the measured distribution resembles the single exponential function [inset of Fig. \ref{fig1}(f)] for all $V_{\mathrm{PS}}$, however, the calculation predicts a double exponential function. Therefore, to achieve the consistency with the measured distribution, at least one of the following conditions must be satisfied:  \textit{``Population of state 5, $C_5^*$ is much smaller than $C_1^*+C_4^*$"} and  \textit{``$\Gamma_{25}^*$ takes a similar value to $\lambda_{-}^*$ ($\sim$ 2 kHz)"}. In fact, we confirmed that both of them hold for our experimental conditions and that the former is the most important.

Using the five-state FCS, first, we analyze $\Gamma_{21}^*$, which is plotted in Fig. \ref{fig2}(a) by open blue circles. Note that because the five-state FCS includes the phonon-induced transitions, the calculation is performed only above the threshold voltage of $V_{\mathrm{PS}}$. $\Gamma_{21}^*$ stays at approximately 2 kHz even with higher $V_{\mathrm{PS}}$, consistent with our expectation of $\Gamma_{21}^*$ being independent of a phonon density. $\Gamma_{21}$ evaluated by the three-state FCS (filled blue circles in Fig. \ref{fig2}(a)) decreases as $V_{\mathrm{PS}}$ increases, but this is because the population of state 1 decreases, so the transition probability from the (0,2) state to state 2 effectively decreases. 

Next, we show the FCS computation results of $\Gamma_{43}^*$ depicted in Fig. \ref{fig3}(b) by open green circles. The rate increases with increasing $V_{\mathrm{PS}}$, because the phonon density at phonon energy $E_{\mathrm{ph}}=\Delta$ rises. The evaluated results are qualitatively reproduced by our model proposed above. To confirm the validity of the evaluated transition rates, we perform the theoretical calculation of $\Gamma_{43}^*$. In this calculation, the electron-phonon interaction originating from the deformation potential and both the Rashba- and Dresselhaus- type spin-orbit interaction (see section VII in Supplemental Material) are considered. The phonon density at $E_{\mathrm{ph}}=\Delta$ is derived from the effective phonon temperature $T_{\mathrm{ph\_ 34}}$, assuming the Boltzmann distribution $\Gamma_{43}^*/\Gamma_{34}^*=\exp \left[-\Delta/(k_BT_{\mathrm{ph\_ 34}})\right]$. Note that $T_{\mathrm{ph\_ 34}}$ is determined by the ratio $\Gamma_{43}^*/\Gamma_{34}^*$ ,which is experimentally evaluated by the five-state FCS as shown in Fig. \ref{fig4}(a).  We calculate the transition rates with $\Delta$ and the inter-dot distance $2d$ as parameters, and find the quantitative agreement with the values obtained by the five-state FCS using $\Delta=0.85\ \mathrm{meV}$ and $2d=260\ \mathrm{nm}$, as depicted in Fig. \ref{fig3}(b). These values of $\Delta$ and $2d$ are comparable to those estimated from the bias triangle (see section III in Supplemental Material) and the designed inter-dot distance ($2d=250$ nm). 

We also theoretically calculate $\Gamma_{41}^*$ using a similarly defined effective phonon temperature $T_{\mathrm{ph\_ 14}}$, given by $\Gamma_{41}^*/(2\Gamma_{14}^*)=\exp \left[-\Delta/(k_BT_{\mathrm{ph\_ 14}})\right]$, using the same values of $d$ and $\Delta$ as the previous paragragh [see also Fig. \ref{fig4}(a) for the ratio $\Gamma_{41}^*/(2\Gamma_{14}^*)$] . Figure \ref{fig3}(c) shows the calculated $\Gamma_{41}^*$, for the same conditions of $\Delta$ and $d$ as $\Gamma_{43}^*$. From this computation, the intra-dot electron relaxation rate $\Gamma_{14}^*$ can also be evaluated as above 20 kHz, which is comparable with the value reported previously \cite{Fujisawa2002, PhysRevLett.95.056803, PhysRevLett.94.196802}.

\section{Discussion}
\begin{figure}[h]
\centerline{ \includegraphics[width=\linewidth]{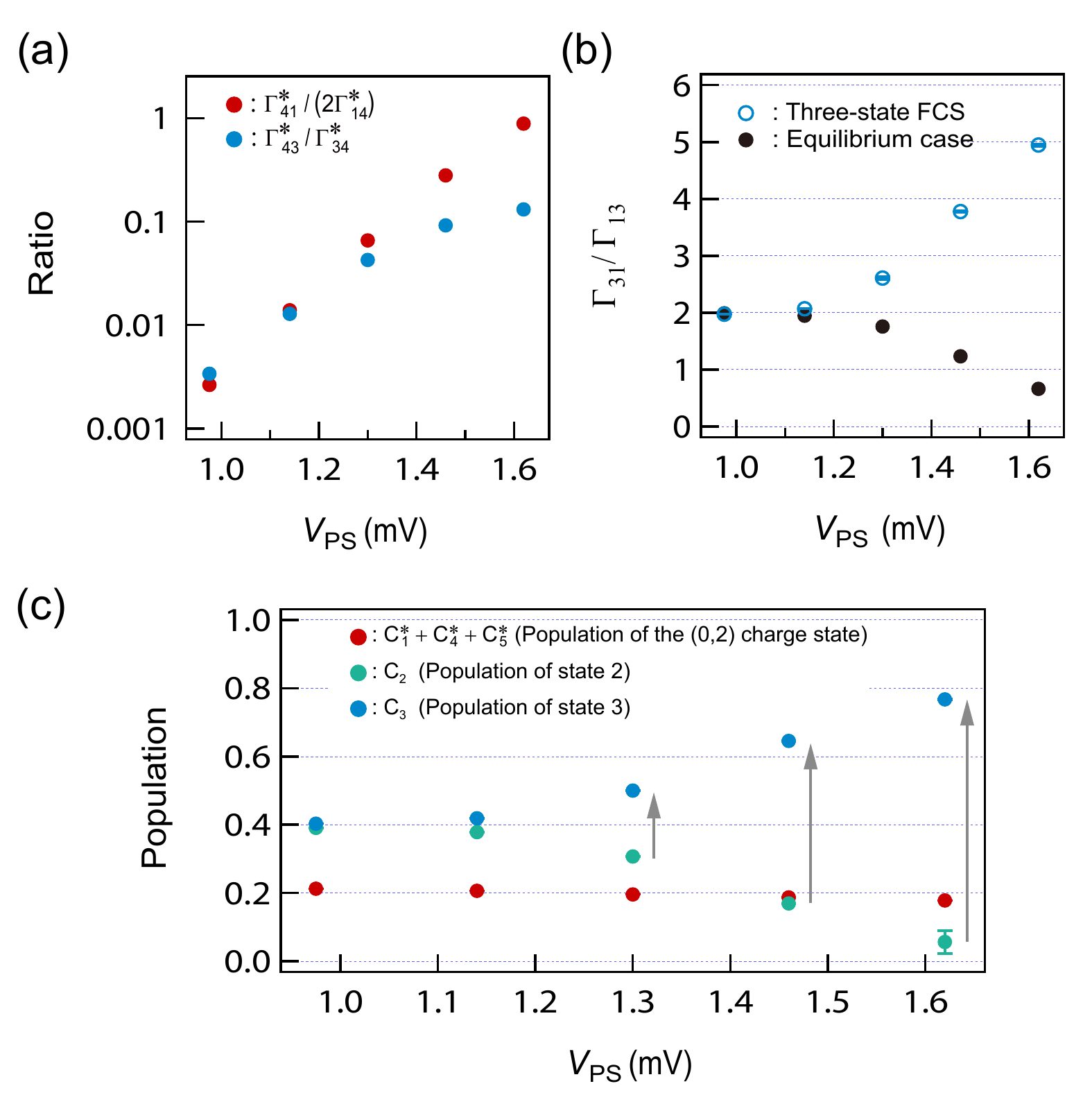} }
\caption{(a) Ratios of intra-dot phonon excitation and relaxation ($\Gamma_{41}^*/(2\Gamma_{14}^*)$, red circles) and of inter-dot phonon excitation and relaxation ($\Gamma_{43}^*/\Gamma_{34}^*$, blue circles), plotted with respect to $V_{\mathrm{PS}}$. The former ratio is more enhanced at higher $V_{\mathrm{PS}}$ because more phonons arrive at the right QD than at the left QD, and the intra-dot excitation occurs more frequently than the inter-dot excitation. (b) Ratio of the total spin-flip tunnel rates $\Gamma_{31}$/$\Gamma_{13}$ versus $V_{\mathrm{PS}}$ evaluated by the three-state FCS (open blue circles). For lower $V_{\mathrm{PS}}$, the ratio is approximately 2. In contrast, when applying higher bias voltage $V_{\mathrm{PS}}$, the ratio increases gradually. For the case in which there is no phonon density gradient between the two dots, $\Gamma_{31}/\Gamma_{13}$ follows the $V_{\mathrm{PS}}$ dependence depicted by filled black circles and decreases to $2/3$. (c) Populations of the (0,2) charge state and states 2 and 3 depicted in Fig. \ref{fig3}(a). For states 2 and 3 of the (1,1) charge states, the population of parallel spin configurations ($C_3$) increases, whereas that of anti-parallel spin configurations ($C_2$) decreases. }
\label{fig4} 
\end{figure}
Finally, we discuss the nonequilibrium properties of the phonon-induced charge-spin dynamics. Because the distances from the phonon source to the left and right QDs are different, the phonon density is different at the two dot positions if the generated phonon is in nonequilibrium. Therefore, this introduces a difference between the ratio of $\Gamma_{14}^*$ to $\Gamma_{41}^*$ and that of $\Gamma_{34}^*$ to $\Gamma_{43}^*$. We plot these ratios in regard to $V_{\mathrm{PS}}$ in Fig. \ref{fig4}(a). It should be noted that we divide the ratio of $\Gamma_{14}^*$ to $\Gamma_{41}^*$ by 2 because there are two available destinations in the transition from $\ket{S(0,2)}$ to $\ket{T_{\pm}(0,2)}$ whereas there is only one in the transition from $\ket{\uparrow\uparrow(1,1)}$ ($\ket{\downarrow\downarrow(1,1)}$) to $\ket{T_{+}(0,2)}$ ($\ket{T_{-}(0,2)}$ ). The former ratio is larger at higher $V_{\mathrm{PS}}$, indicating that more acoustic phonons would arrive at the right QD than at the left QD, thus inducing more frequent excitations of electron states in the right QD. Moreover, these ratios determine the abovementioned effective phonon temperatures $T_{\mathrm{ph\_ 34}}$ and $T_{\mathrm{ph\_ 14}}$, and the latter temperature at the right QD increases more. This indicates that the lattice temperature gradient would be created between the two dots of the DQD. 

For further discussion of the nonequilibrium property of the spin-flip tunnel process induced by the phonon density gradient, we consider the spin-flip tunnel ratio $\Gamma_{31}/\Gamma_{13}$. By comparing the distributions of the three- and five-state FCS, $\Gamma_{13}$ and $\Gamma_{31}$ in the three-state FCS are the total spin-flip tunnel rates between state 3 and the (0,2) charge state in Fig. \ref{fig3}(a).
\begin{align*}
\Gamma_{13}&=\Gamma_{13}^*+\Gamma_{43}^*,\\
\Gamma_{31}&\simeq\frac{\Gamma_{14}^*}{\Gamma_{14}^*+\Gamma_{41}^*}\Gamma_{31}^*+\frac{\Gamma_{41}^*}{\Gamma_{14}^*+\Gamma_{41}^*}\Gamma_{34}^*,
\end{align*}
where the second equation is derived from the approximation of $\Gamma_{14}^*+\Gamma_{41}^*$ being much larger than the other rates (see Section VI B in Supplementary Material). Moreover, to be precise, the second term of $\Gamma_{13}$ is the phonon-assisted inter-dot tunnel without spin flip, but the fast spin relaxation $\Gamma_{14}^*$ follows it. Therefore, $\Gamma_{43}^*$ is mostly equal to the phonon-induced spin-flip tunnel rate from state 3 to 1. The ratio $\Gamma_{31}/\Gamma_{13}$ plotted as open blue circles in Fig. \ref{fig4}(b) shows the increase with $V_{\mathrm{PS}}$ above 1.1 mV, similar to that for $\Gamma_{13}$ in Fig. \ref{fig2}(a). To prove that the increase in the ratio is not due to the phonon excitation but to the spatial unbalance of the phonon density, we consider the equilibrium condition, i.e., the case in which the two ratios $\Gamma_{41}^*/(2\Gamma_{14}^*)$ and $\Gamma_{43}^*/\Gamma_{34}^*$ are the same. For this equilibrium case, the ratio of the spin-flip rates is simplified to $\Gamma_{31}/\Gamma_{13}\simeq2\Gamma_{41}^*/(\Gamma_{14}^*+\Gamma_{41}^*)$, where we assume that the spin-flip tunnel rates in the ground states are not affected by the phonon density gradient, i.e., the relation $\Gamma^*_{31}=2 \Gamma^*_{13}$ is retained even for $V_{\mathrm{PS}}>0$ mV. The ratio $\Gamma_{31}/\Gamma_{13}$ at the equilibrium condition is shown by filled black circles in Fig. \ref{fig4}(b). For the higher phonon density above the threshold, the ratio starts to decrease from 2 to 2/3, in contrast to our experimental result. This $V_{\mathrm{PS}}$ dependence indicates that the ratio of the total spin-flip tunnel rates is strongly influenced by the spatial gradient of the phonon density. This non-equilibrium spin-flip tunneling results in a pumping-like effect for the (1,1) charge state. In Fig. \ref{fig4}(c), we show the populations of the (0,2) charge state ($C^*_1+C^*_4+C^*_5$) and states 2 and 3 in regard to $V_{\mathrm{PS}}$. At lower $V_{\mathrm{PS}}$, these populations are approximately 0.2, 0.4, and 0.4, respectively. In this case, as the populations are dominated by states 1, 2, and 3, and their energies are equal, these population values are determined by the number of available internal states in states 1, 2, and 3. For higher $V_{\mathrm{PS}}$, however, the population of state 3 increases from 0.4, whereas that of state 2 decreases. In contrast, the population of the (0,2) charge state remains at 0.2. This indicates that the population of the (1,1) charge state is transferred from state 2 to 3, and the parallel spin configuration becomes more probable. 

In conclusion, we study the spin-charge cooperative dynamics in a DQD under a nonequilibrium phonon environment.
The tunnel rates for the spin-flip and spin-conserving processes in the DQD with a side contact QD as a phonon source are evaluated using the FCS technique. The spin-flip tunnel rate is significantly enhanced
above a certain threshold of applied bias voltage $V_{\mathrm{PS}}$ on the phonon source QD that determines the maximum energy of the generated acoustic phonons. We propose a mechanism of the spin-flip process intermediated by $\ket{T_{\pm}(0,2)}$ with phonon excitation. The extended-FCS model evaluates the phonon-induced transition rates and find quantitative consistency with the predictions from our theoretical calculations, supporting the validity of our FCS model. Finally, we show that the spatial gradient of the phonon density between the two dots of the DQD by computing the ratios of the phonon excitation rate to the relaxation rate as depicted in Fig. \ref{fig3}(a), indicating the local temperature gradient over the DQD would be created. Further, the ratio of the total spin-flip tunnel rates $\Gamma_{31}/\Gamma_{13}$ increases from the equilibrium value at higher $V_{\mathrm{PS}}$. We prove that the asymmetric enhancement of the two spin-flip tunnel rates originates from the spatial gradient of the phonon density, by comparing it with the result under the equilibrium phonon condition with no bias voltage. The result indicates that the spin-charge dynamics are strongly affected by the nonequilibrium phonon distribution and the populations of the spin states are strongly modified. Our findings may promote new concepts of DQD heat engines and thermoelectric devices that are driven and controlled by a local lattice temperature gradient.

\begin{acknowledgments}
 This work was supported by a Grant-in-Aid for Young Scientific Research (A) (No. JP15H05407), Grant-in-Aid for Scientific Research (A) (No. JP16H02204, No. JP25246005), Grant-in-Aid for Scientific Research (S) (No. JP26220710, No. JP17H06120), Grant-in-Aid for Scientific Research (B) (No. JP18H01813) from Japan Society for the Promotion
of Science, JSPS Research Fellowship for Young Scientists (No. JP16J03037), JSPS Program for Leading Graduate Schools (ALPS) from JSPS, Japan Society for the Promotion of Science (JSPS) Postdoctoral Fellowship for Research Abroad Grant-in-Aid for Scientific Research on Innovative Area, "Nano Spin Conversion Science" (No. 26103004), and Grant-in-Aid for Scientific Research on Innovative Area, CREST (No. JPMJCR15N2). A.D.W., A.L., and S.R.V gratefully acknowledge the support of DFG-TRR160, BMBF-Q.com-H 16KIS0109, and the DHF/UFA CDFA-05-06. Y.T. is supported by a Grant-in-Aid for Scientific Research (C) (No. JP18K03479). 
\end{acknowledgments}

%
\end{document}